\documentclass[useAMS,usenatbib]{mn2e}

\usepackage{mathptmx}
\usepackage[T1]{fontenc}
\usepackage{ae,aecompl}
\usepackage{lscape}
\usepackage{graphicx}	% Including figure files
\usepackage{amsmath}	% Advanced maths commands
\usepackage{amssymb}	% Extra maths symbols
\usepackage{times,graphicx,amsmath,amsfonts,amssymb,epstopdf}
\usepackage[normalem]{ulem}

\title[Spectropolarimetry of Herbig Ae/Be Stars]{A Statistical Spectropolarimetric Study of Herbig Ae/Be Stars}

\author[K. M. Ababakr, R. D. Oudmaijer and J.S. Vink]{K. M. Ababakr$^{1}$\thanks{E-mail:
K.M.Ababakr@leeds.ac.uk}, R. D. Oudmaijer$^{1}$ and J.S. Vink$^{2}$\\
  $^{1}$School of Physics and Astronomy, University of Leeds, EC Stoner Building, Leeds LS2 9JT, UK\\
$^{2}$Armagh Observatory, College Hill,  Armagh BT61 9DG, UK\\}

\date{Accepted XXX. Received YYY; in original form ZZZ}

\pubyear{2015}

\begin{document}
\label{firstpage}
\pagerange{\pageref{firstpage}--\pageref{lastpage}}
\maketitle

\begin{abstract}

We present H$\alpha$ linear spectropolarimetry of a large sample of
Herbig Ae/Be stars. Together with newly obtained data for 17 objects,
the sample contains 56 objects, the largest such sample to date. A
change in linear polarization across the H$\alpha$ line is detected in
42 (75 $\%$) objects, which confirms the previous finding that the
circumstellar environment around these stars on small spatial scales
has an asymmetric structure, which is typically identified with a
disk. A second outcome of this research is that we confirm that Herbig
Ae stars are similar to T Tauri stars in displaying a line polarization
effect, while depolarization is more common among Herbig Be
stars. This finding had been suggested previously to indicate that
Herbig Ae stars form in the same manner than T Tauri stars through
magnetospheric accretion. It appears that the transition between these
two differing polarization line effects occurs around the B7-B8 spectral type.
This would in turn not only suggest that Herbig Ae stars accrete in a
similar fashion as lower mass stars, but also that this accretion
mechanism switches to a different type of accretion for Herbig Be stars.
We report that the magnitude of the line effect caused by
electron scattering close to the stars does not exceed 2\%. Only a
very weak correlation is found between the magnitude of the line
effect and the spectral type or the strength of the H$\alpha$
line. This indicates that the detection of a line effect only relies
on the geometry of the line-forming region and the geometry of the
scattering electrons.

\end{abstract}


%%%%%%%%%%%%%%%%%%%%%%%%%%%%%%%%%%%%%%%%%%%%%%%%%%

%%%%%%%%%%%%%%%%% BODY OF PAPER %%%%%%%%%%%%%%%%%%

\section{Introduction}

Herbig Ae/Be (HAeBe) stars, the more massive counterparts of T Tauri
stars, are optically visible pre-main-sequence (PMS) stars with masses
roughly between 2 and 10~M$_{\odot}$. This group of stars was first
identified by \citet{1960ApJS....4..337H}.  With their intermediate
masses, they play an essential role in addressing the formation of
high mass stars as they bridge the gap between low mass stars, whose
formation is fairly well understood and high mass stars, whose
formation still poses challenges.  Lower mass stars are thought to
form through magnetically controlled accretion (MA,
e.g. \citet{2007bouvier}, whereas evidence for this mode of accretion
is lacking for high mass stars. Not only are higher mass stars not
expected to be magnetic as their radiative envelopes would inhibit the
presence of the magnetic fields that dominate the accretion in low
mass objects, these magnetic fields have also hardly been detected
(e.g. \citealt{alecian13}). Traditionally, the change from MA and a
different, hitherto unexplored, accretion mechanism was thought to be
around the spectral type boundary M/K to F/A where the envelope
structure changes from convective to fully radiative. However, it had
long been known that Herbig Ae stars have different properties than
Herbig Be stars and may form in a different manner
(e.g. \citealt{fuente1998, 1999testi,ilee2014} who studied millimeter
emission, clustering properties and CO first overtone emission
properties respectively). In addition, from various recent studies it
appears that Herbig Ae stars are more similar to the T Tauri stars
than to Herbig Be stars. For example, \citet{grady2010} infer that the
accretion shock regions in a Herbig Ae star are comparable in size and
location to those in the magnetic lower mass objects.
\citet{scholler2016} interpret from the observed spectroscopic
variability that a Herbig Ae star is currently undergoing
magnetically-controlled accretion in the same manner as the T Tauri
stars. In contrast, dedicated modeling indicated that, if present at
all, the magnetosphere in a Herbig Be star with spectral type B9IV
must be small \citep{kurosawa2016}. Indeed, \citet{patel2017} find
that magnetic fields are not required to explain the spectroscopic
properties of early type Herbig Be stars. Finally, interferometric
studies show that some of the hotter Herbig Be stars have much smaller
near-infrared sizes than would be expected from the dust sublimation
radius. This can be explained by optically thick gas in accretion
disks reaching to the star \citep{kraus2008, kraus2015}. A first
indication that Herbig Ae stars share similarities with T Tauri stars
but are different from Herbig Be objects emerged from a study into
linear spectropolarimetry across H$\alpha$ emission by
\citet{Vink2002, vink03}.  Later, these authors found that the linear
spectropolarimetric signatures observed around the H$\alpha$ emission
line in both T Tauri and Herbig Ae stars can be explained by compact
H$\alpha$ emission scattered off a circumstellar disk
(\citealt{vink05a}, Vink, Harries \& Drew 2005,
\citealt{Mottram07}). This is suggestive of the notion that Herbig Ae
stars may form in the same manner as T Tauri stars, where the compact
H$\alpha$ emission could arise from accretion hot spots or funnels due
to magnetospheric accretion.

  Further clues to this effect were presented by \citet{cauley2015}
  who found that the emission and absorption line properties of Herbig
  Be stars are significantly different from Herbig Ae stars, who in
  turn seem to have properties intermediate between Herbig Be and T
  Tauri stars. Finally, \citet{fairlamb15} found that the UV-excess of
  Herbig Ae stars can be explained by magnetospheric accretion, but
  that the earliest Herbig Be stars have too large UV-excesses to be
  explained by the usual accretion shock scenario
  \citep{Muzerolle04}. For recent reviews on the subject of accretion
  in young stellar objects, including Herbig Ae/Be stars, we refer the
  reader to \citet{hartmann2016}, \citet{beltran2016} and
  \citet{rene2017}.

Linear spectropolarimetry is an effective technique to probe ionised
inner circumstellar disks around stars on scales of order stellar
radii, scales that are small enough to probe the accretion region of
young stars.  The technique was first successfully used to probe the
circumstellar disks around classical Be stars
\citep{1974MNRAS.167P..27C, 1976ApJ...206..182P}.  These authors
  demonstrated that the continuum light is scattered and polarized by
  free electrons, while the hydrogen recombination line emission,
  which arises from a volume larger than where the electron scattering
  dominates, is not or hardly polarized. The use of the technique was
extended by \citet{1999MNRAS.305..166O}, \citet{Vink2002} and
\citet{Mottram07} to cover HAeBe stars.  \citet{Vink2002} classified
HAeBe stars according to their spectropolarimetric signature. In a
study of 23 HAeBe objects, they found that many of the HBe stars (7
out of 12) show a depolarisation line effect consistent with a
circumstellar disk, similar to that observed in Be stars, while an
intrinsic polarisation line effect, as also observed in T Tauri stars
is more dominant in most, 9 out of 11, less massive HAe stars.
Their results suggest a physical switch from line polarisation for HAe
stars to depolarisation in HBe stars.

To put these results on a firmer footing, a larger sample is needed to
support and confirm the findings and draw statistical conclusions.  In
this work we aim to provide a statistical investigation into the
spectropolarimetric properties of HAeBe stars and their relation with
their lower mass counterpart T Tauri stars. By collating all the data
in the literature and adding newly obtained data, we present the
spectropolarimetric results of a sample of 56 HAeBe objects, nearly
three times larger than the sample of \citet{Vink2002}.  The paper is
structured as follows. In Section 2, in order to properly interpret
the data of this large sample, we start by an updated overview of the
use of linear spectropolarimetry. This is followed by a discussion of
the details of the sample selection, the complementary observations
and the data reduction. In Section 3, we present the results which are
discussed in Section 4. Finally, we conclude in Section 5.

\section{Methodology, observations and data reduction}

\subsection{Spectropolarimetry as a probe of inner regions}

The technique of linear spectropolarimetry as applied here, exploits
the fact that radiation will be scattered by free electrons in an
ionized region. This results in the light to be polarized in the plane
of the sky perpendicular to the original direction of travel. In case
of a projected circular geometry of the scatterers on the sky - for
example a disk observed pole-on or a spherical distribution - all
polarization vectors cancel, resulting in a net zero
polarization. However, in case of an asymmetric distribution, a net
polarization is observed.  Most of the polarization due to free
electrons is found to occur within a few stellar radii and results in
polarizations of order 1-2\% \citep{cassinelli1987}.  We can take
advantage of the fact that the stellar continuum photons and emission
line photons originate from different locations and scatter
differently off the free electrons, resulting in a different
polarization across line and continuum. The spectropolarimetry around
an emission line, most often H$\alpha$ has been observed, can
therefore probe scales very close to the star.  We note that many
other scattering agents can give rise to polarization such as dust.
However, whether circumstellar or interstellar, dust is located far
away from the stellar and line emitting regions and it has a very
broad wavelength dependence. As a consequence, the continuum and line
radiation will be scattered in a similar fashion and no line-effect is
visible (e.g. \citealt{trammell94}).

\begin{figure*}
       \centering
       \includegraphics[width=16cm]{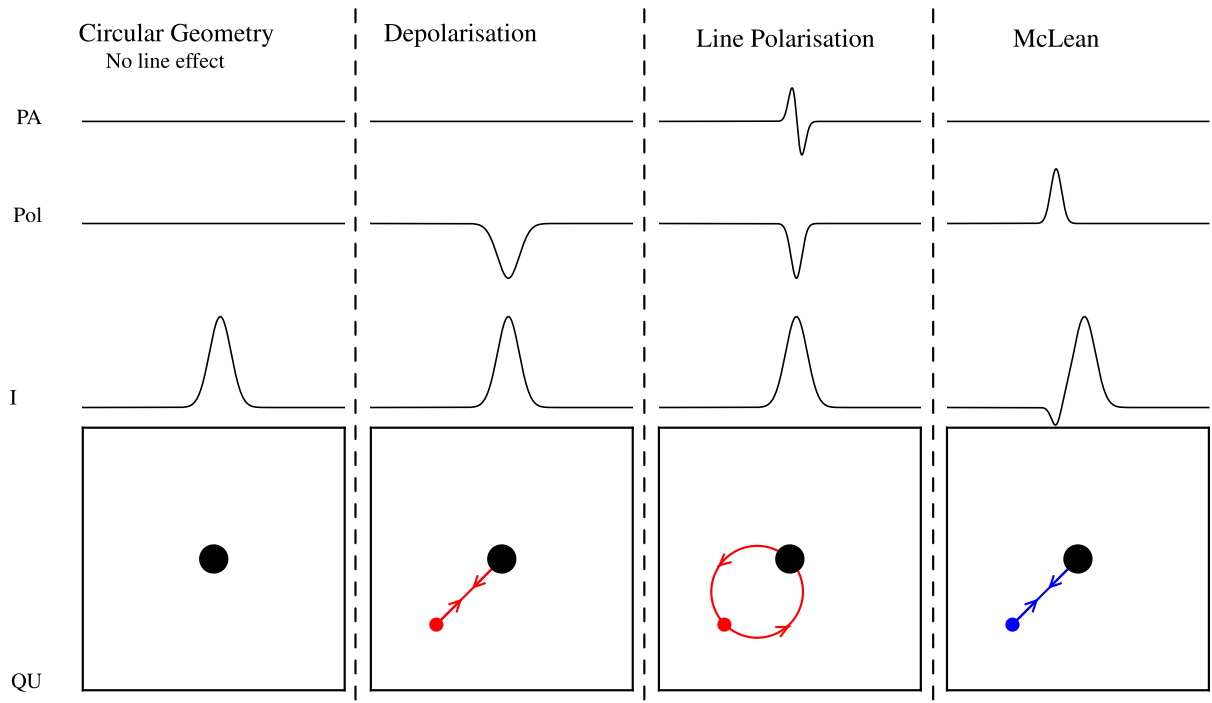}
   \caption[]{Schematic showing spectropolarimetry expectations across the H$\alpha$ line in triplots (top) and {(\it Q, U)} diagrams
     (bottom). In the triplot, the Stokes
     intensity (I) is shown in the bottom panel, polarisation (\%) in
     the centre, while the position angle (PA) is shown in the upper
     panel. The first column shows where the geometry of the circumstellar environment on the sky is circular and hence no line effect is detected. The other three columns show where the geometry on the sky is not circular and a line effect is seen. The second column shows a depolarisation line effect, note that the depolarisation across H$\alpha$ line is as broad as Stokes I. This depolarisation line effect translates into {(\it Q, U)} diagram as a linear excursion from continuum knot towards the central line. The arrows in the {(\it Q, U)} diagrams indicate the polarisation moves in and out of the line effect from blue to red wavelengths. Intrinsic line polarisation is shown in third column, where H$\alpha$ line from the accretion compact region is scattered in a rotating disk. In this case the polarisation across H$\alpha$ is narrower that the width of Stokes I and a flip is seen in PA caused by a rotating disk. This flip is seen as a loop in {(\it Q, U)} diagram. Finally column four shows a different polarisation signature across the absorption component of H$\alpha$ which is commonly known as the McLean effect. The figure is adapted from \citet{Vink2002}.}
    \label{cartoon}
\end{figure*}

Thus far, in the study of Herbig Ae/Be stars, three types of
``line-effect'' can be identified and whereas the nature of the
electron scattering process may be different, they share the fact that
they probe the small scales of the electron scattering region.  Let us
start with the so-called depolarization effect. This had already been
observed toward Be stars in the early seventies, and was in fact the
first (indirect) proof that these stars are surrounded by small disks.
The density of free electrons is the highest closest to the star,
which is where the continuum photons emitted by the stellar
photosphere will be polarized.  In contrast, the hydrogen
recombination line emission will be less polarised as it passes
through a smaller ionised volume and hence encounters fewer free
electrons.  Therefore, a detection of a depolarization line effect
across an emission line can immediately trace the presence of an
ionised asymmetric structure with a size of order stellar radii. This
is on much smaller smaller scales than can be typically probed by the
best imaging techniques
(e.g. \citealt{kraus15,mendigutia15,2017mendigutia} on
optical/near-infrared interferometry of Herbig Ae/Be stars). Such a
depolarization has been identified and confirmed to be due to a disk,
as the polarization angles indicate that the scattering region is
parallel to disks that have been observed at larger scales
(e.g. \citealt{quirrenbach1997} for classical Be stars, and
\citealt{wheelwright11} for Herbig Ae/Be stars, see reviews by
\citealt{oudmaijer07} and \citealt{vink2015}).  This depolarisation
line effect, similar to that observed in classical Be stars, follows
the emission line and is more or less as broad as the emission. It can
be detected in the polarisation spectra, but also in the position
angle spectra or both, depending on the (vector) contribution of the
interstellar polarization. When translated into the Stokes {\it QU}
parameters and plotted in the {\it (Q, U)} diagram, we will see a line
excursion from the continuum towards the line (see Fig. \ref{cartoon},
second column).

Alternatively, a different line effect signature can be detected when
an emission line emerges from a compact, central, region and scatters
off a circumstellar disk.  This is usually accompanied by a flip in
polarisation or polarisation angle (Vink, Harries \& Drew 2005). When
mapped on the {\it (Q, U)} diagram, this particular line effect appears as a
loop across the emission line (see Fig. \ref{cartoon}, third
column). The compact region is thought to be due to magnetospheric
accretion, where the material from the disk is funnelled via accretion
columns on to the stars, causing a shock on the photosphere. This type
of line effect has been observed in T Tauri stars and HAe stars
\citep{vink05a}.  Importantly, the data provide the position angles of
the disks in agreement with previous observations, while the models
reproducing the line effect return disk sizes of order stellar radii,
also in agreement with observations (Vink, Harries
\& Drew 2005, \citealt{vink2015}).

Thirdly, the absorption component of the emission line also produces a
line effect signature that is commonly referred to as the McLean
effect \citep{McLean1979}.  In this case, an enhanced polarisation is
detected across the absorption accompanying an emission line as
direct, unscattered, light from the star is absorbed.  This normally
results in observing a typical (inverse) P Cygni line profile,
depending on whether we consider infall of material or an outflow
respectively. The absorbed photons will be re-emitted isotropically
and part of the emission will be scattered into our line of sight.
%This is why the flux in the absorption often does not reach zero.
If the distribution of the, inner, scattering material is not circular on the
sky, an enhanced polarisation will be detected across the absorption
compared with the continuum light (see Fig. \ref{cartoon}, fourth
column).

Last but not least, if the geometry of the region containing the
scatterers is circular on the sky, no line effect would be
detected in most of the above cases (see Fig. \ref{cartoon}, first
column) since all polarisation vectors will cancel. Such a circular
geometry could be due to a spherical distribution of material or if
the disk is viewed pole-on\footnote{A possible exception to this
  statement can occur in the case of the intrinsic line
  polarization. Emission emerging from an anisotropic source,
  such as an accretion hot spot, scattering off a circular geometry
  would also result in net polarization. However, we note that the
  position angles and size scales derived from our data are consistent
  with circumstellar disks in the case of the T Tauri stars.}

\begin{table*}

\caption[Log of spectropolarimetric observations of the ISIS
  sample]{New HAeBe observations. Columns 1 \& 2 show the target
  names, RA and Dec. are tabulated in columns 3 \& 4.  The $\it V$
  band magnitude and spectral type are taken from
  \citet{1994A&AS..104..315T} and SIMBAD, while for objects with new
  determinations of the photospheric temperature \citep{alecian13,
    fairlamb15}, the temperatures have been converted into a spectral
  type using Schmidt-Kaler's tables in \citet[Chap. 4]{landolt1982}. These
  are listed in column 5 \& 6. The integration times (column 8) denote
  the total exposures. Column 9 gives the SNR.}

\begin{tabular}{l l l l r r r r r }
\hline
Name   & Alt. name&RA (J2000) & Dec.(J2000)&   {\it V}&Spec. type& Obs date  &Exposure (s)&SNR\\

\hline 
HD 163296   & MWC 275 & 17:56:21.3&-21:57:21.9&    6.9   &    A1       & 04-08-15     &12$\times$20&800\\ %herbig cata both%fairlamb spect
MWC 610  & HD 174571 & 18:50:47.2  &+08 42 10.1&  8.9   &    B2       & 04-08-15     & 4$\times$240&470    \\%s cauley m simbad
MWC 342   & V1972 Cyg & 20:23:03.6&+39 29 50.1&    10.6   &    B(e)       & 04-08-15     &24$\times$20&275   \\ %simbad both
HD 200775 &  MWC 361 &21:01:36.9 &+68:09:47.8&   7.4   &    B3       & 04-08-15     & 24$\times$15&860   \\ %s herbig m simbad
HD 203024  &BD+68 1195  & 21:16:03.0&+68:54:52.1&    8.8   &    A1       & 04-08-15     &4$\times$240&480  \\ %herbig both
V361 Cep   & AS 475  &21:42:50.2 &+66:06:35.1&    10.2   &    B3       &   04-08-15  &8$\times$300&660 \\%s cauley m simbad
HD 240010 &MWC 655 & 22:38:31.8 &+55:50:05.4&   9.2   &    B1       &  04-08-15  &   4$\times$240&545   \\ %simbad both
V374 Cep   & AS 505 &23:05:07.6 &+62:15:36.5&    10.6   &    B0       &   04-08-15   &8$\times$300&400  \\%herbig both
MWC 863   & HD 150193& 16:40:17.9&-23:53:45.2&    8.8   &    A2       & 05-08-15     &12$\times$150&780  \\ %herbig both
V718 Sco  & HD 145718 &16:13:11.6 &-22:29:06.7&    8.9   &   A7       & 05-08-15     & 12$\times$120&740   \\%s cauley m simbad
HD 141569  & PDS 398  &15:49:57.7 &-03:55:16.3&    7.1   &    A0       & 05-08-15     &8$\times$90&725   \\ %s herbig m simbad
MWC 300   & 	HBC 283  &18:29:25.7 &-06:04:37.3&   11.6   &    B1       & 05-08-15     & 20$\times$90&260    \\% s herbig m simbad
MWC 953   &ALS 9906 &18:43:28.4 &-03:46:16.9&    10.8   &    B2       & 05-08-15     &12$\times$180&370  \\ %simbad both
MWC 623    & V2028 Cyg & 19:56:31.5&+31:06:20.1&    10.9   &    B4(e)       &   05-08-15  &24$\times$45&90 \\ %simbad both
V1977 Cyg &AS 442 & 20:47:37.5&+43:47:24.9&    10.9   &    B8       &  05-08-15  &   12$\times$300,4$\times$45&475      \\% s herbig m simbad
MWC 1080   & V628 Cas &23:17:25.6&+60:50:43.4&     11.6   &    B0       &   05-08-15   &20$\times$300&440 \\ %s herbig m simbad
HD 144432   &  PDS 78 &16:06:57.9&-27:43:09.8&     8.2   &   A8I       &   05-08-15   &12$\times$150&680   \\% s herbig m simbad
\hline
\label{obser}
\end{tabular}
\end{table*}

\subsection{Construction of the sample}

In order to perform a statistical study on the line polarimetry, a
large sample of HAeBe stars is needed. We combined all our previous
spectropolarimetric work across H$\alpha$ line of HAeBe stars
\citep{1999MNRAS.305..166O,
  Vink2002,vink05a,Mottram07,wheelwright11} into one
sample. Our previous medium resolution linear spectropolarimetric data
were obtained using the RGO spectrograph on the 3.9-m Anglo Australian
Telescope (AAT) \citep{1999MNRAS.305..166O}, the ISIS spectrograph on
the 4.2-m on the William Herschel Telescope (WHT), La Palma
\citep{1999MNRAS.305..166O,
  Vink2002,vink05a,Mottram07,wheelwright11,ababakr16} and the FORS2
spectrograph mounted on ESO's 8.2-m Very Large Telescope (VLT) in
Chile \citep{ababakr16}.  These bring the total number of observed
HAeBe objects to 56 (31 HBe and 25 HAe). These objects are presented
in Table \ref{sss}. A sample of 29 HAeBe stars, of which 19 objects
are in common with our sample, was observed spectropolarimetrically by
\citet{harrington09} but due to technical issues, these authors did
not have information on the polarisation angle, and we decided not to
include the remaining 10 objects for the analysis.

Our sample is the largest (linear) spectropolarimetric survey of HAeBe stars
that has been published to date. The vast majority, 52, of the objects
were selected from the HAeBe catalogue of
\citet{1994A&AS..104..315T}, 10 of which are in other tables (extreme
emission lines, other early emission line stars and non-emission line
early type stars) in \citealt{1994A&AS..104..315T}). We proceed under the
assumption that they are young stars. The remaining 5 objects were
taken from the HAeBe candidate stars of
\citet{2003AJ....126.2971V}.  The final sample covers nearly 50\%
of the HAeBe catalogue, where the majority was chosen from the
northern hemisphere. Most of the remaining targets in the catalogue are
too faint (V$\geq$13.5) and would require very long exposure times
as the spectropolarimetry needs high SNR. The combined H$\alpha$
spectropolarimetric observations allow us to conduct the most powerful
statistical investigation into the nature of linear polarisation in the
circumstellar environment of HAeBe stars. To compare the
spectropolarimetric results of HAeBe stars and their lower mass
counterparts T Tauri stars, the spectropolarimetric results of a
sample of 9 T Tauri stars are taken from \citet{vink05a}.

\subsection{Complementary Observations}

Seventeen targets were selected from the HAeBe catalogue of
\citet{1994A&AS..104..315T} and candidates of
\citet{2003AJ....126.2971V} to complement previous spectropolarimetric
results. The list of objects and the log of the observations are
presented in Table \ref{obser}. The SNR is measured over a range of
10~\rm\AA~ around 6700~\rm\AA~ where the continuum is the flattest and
the spectral lines are absent.

The new linear spectropolarimetric data were obtained with the ISIS
spectrograph on the WHT, La Palma, during the nights of 2015 August 4
and 5. The log of the observations is provided in Table
\ref{obser}. The 1200R grating centred at 6800~\rm\AA, with a spectral
range of 1000~\rm\AA, was employed with a windowed $351\times4200$
pixel CCD and a slit width of 1.0 arcsec. This setup provides a
spectral resolution of $\sim$35 kms$^{-1}$ as measured from arc lines
around the H$\alpha$ line. The seeing was less than 1.0 arcsec
throughout both nights.  The polarisation optics, which consist of a
rotating half-wave plate and a calcite block, were used in order to
perform linear polarisation observations. The calcite block separates
the light into two perpendicularly polarised light beams, the ordinary
(O) and extraordinary (E) beam. One complete set of observations
consists of four exposures with the half-wave plate set at angles:
$10^{\circ}$, $55^{\circ}$, $32.5^{\circ}$, and $77.5^{\circ}$. The
dekker with 18 arcsec slot separation was used to observe the object
and the sky simultaneously. Several cycles of observations per object
were obtained at the four position angles to check for the consistency
of the results. Several short exposures were taken for objects with
strong H$\alpha$ line to avoid saturation. Polarised standard stars
and zero-polarised standard stars were observed each night to
calibrate for the instrumental polarisation and angle offset.
 
The data reduction was carried out using {\sc iraf} \citep{Tody93},
which includes bias subtraction, flat fielding, sky subtraction and
extraction of the O and E spectra. The extracted spectra were imported
into the {\sc tsp} package \citep{1997StaUN..66.....B} to compute the
Stokes parameters. The wavelength calibration was performed using {\sc
  figaro}. For analysis purposes, the data were imported into the {\sc
  polmap} package \citep{Harries96}.  Multiple observations of the
same targets provided a very consistent results. As the observations
were obtained at the parallactic angle to achieve high SNR, the angle
calibration was performed using the observed polarised standard
stars. The instrumental polarisation is found to be $\sim$0.1\% while
the angle offset is found to be less than $0.5^{\circ}$ from the
observation of unpolarised and polarised standard stars. As the
instrumental and interstellar polarisation add a wavelength
independent vector to the observed spectra, we did not correct the
observed polarisation for them.

\begin{figure*}
        
        \centering
       \includegraphics[width=17.75cm]{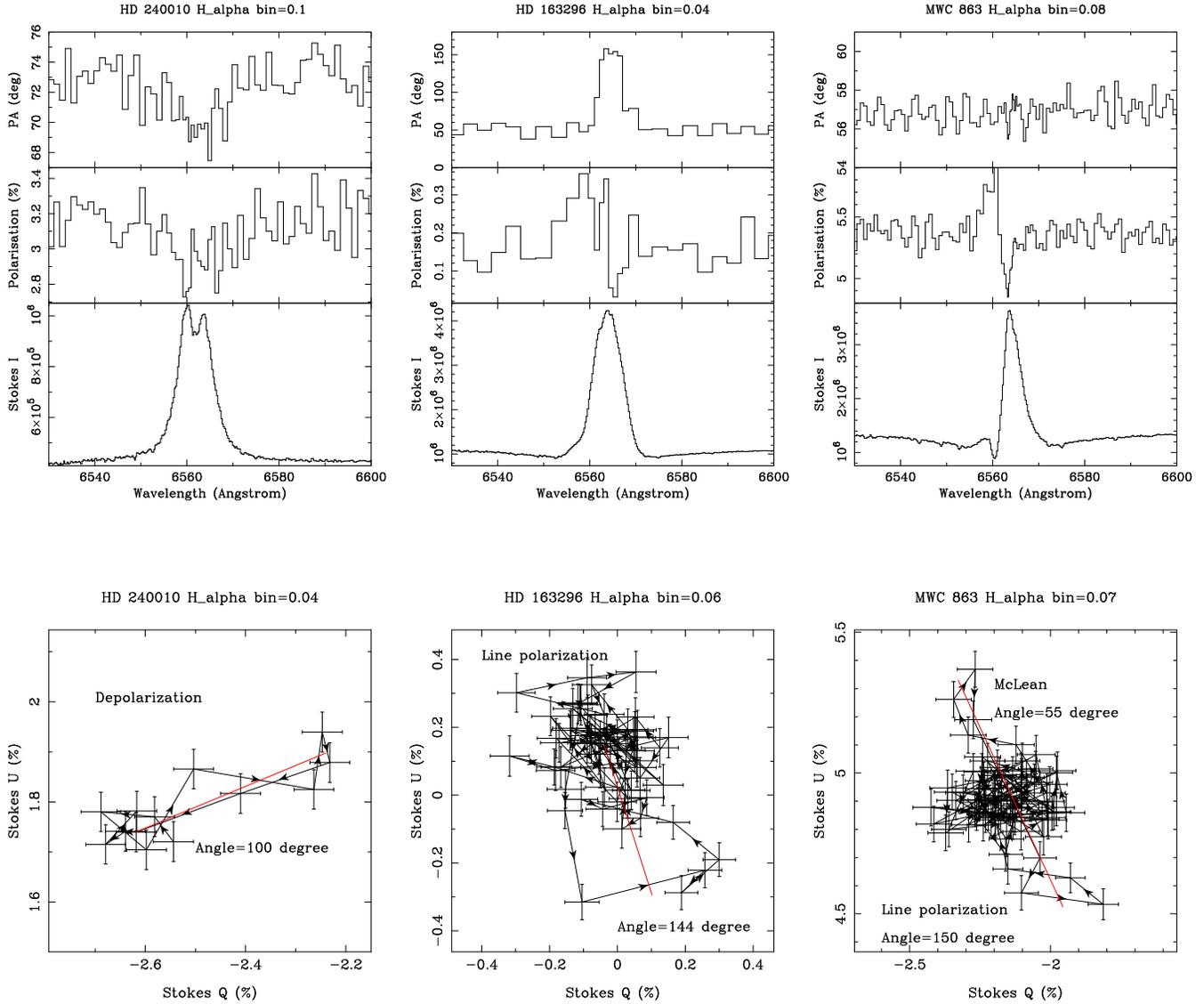}
   \caption{Examples of the various line effect in the H$\alpha$ spectropolarimetry of three objects for which new data were obtained. The data are
     presented as a combination of triplots (top) and {(\it Q, U)} diagrams
     (bottom). In the triplot polarisation spectra, the Stokes
     intensity (I) is shown in the bottom panel, polarisation (\%) in
     the centre, while the position angle (PA) is shown in the upper
     panel. The Q and U Stokes parameters are plotted against each
     other below each triplot. The data are rebinned to a constant
     error in polarisation, which is indicated at the top of each
     plot. The arrows in the {(\it Q, U)} diagrams indicate the polarisation moves in and out of the line effect from blue to red wavelengths. The solid line in the {(\it Q, U)} diagrams represents the direction of the intrinsic polarisation angle. }
    \label{karim}
\end{figure*}

\section{Results}

We begin with a brief presentation of the new data, before we
focus on the statistical results of the full sample of
spectropolarimetric data on HAeBe stars.

\subsection{H$\alpha$ Spectropolarimetry-new observations}

In the new observations 11 objects out of 17 had never been observed
with linear spectropolarimetry.
H$\alpha$ spectropolarimetry was performed for all the targets in
Table \ref{obser} and the results of the entire sample are presented
in Table \ref{sss} and in the Appendix in Fig. \ref{halphafig}.  In total,
11 objects show a possible line effect across the H$\alpha$ line.

\begin{figure*}
        \centering
       \includegraphics[width=\textwidth]{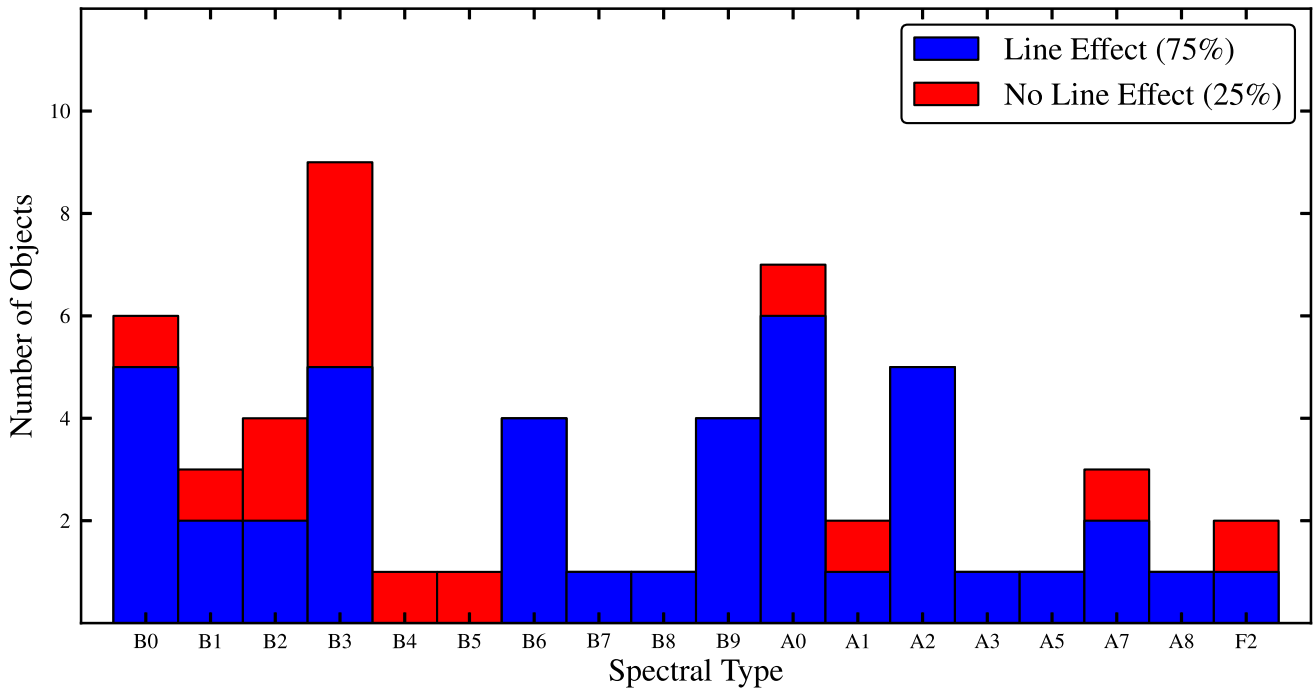}
        
   \caption[Line effect versus spectral type]{The figure shows the
     observed line effect across H$\alpha$ line in each sub group of
     spectral type of the whole sample of HAeBe stars that have been
     observed spectropolarimetrically. Please note that spectral types
     such as A4, A6 and A9 are not always defined in spectral type
     classification schemes, and will thus be missing in a graph such
     as this \citep{jaschek1990}.}
     \label{line_effect}
\end{figure*}

The spectropolarimetric results of three objects from these new
observations, selected as they exhibit the three different line effect
signatures, are shown in triplots in the upper half of
Fig. \ref{karim}. In this triplot, the Stokes I (normal intensity) is
shown in the lower panel, the polarization percentage in the middle,
while the position angle (PA) is displayed in the upper panel. The
results are also represented in a Stokes {(\it Q, U)} diagram (bottom)
in Fig. \ref{karim} using the same wavelength range of the triplot
spectra, but sometimes with a different binning.  HD 240010 shows a
broad depolarisation line effect across the H$\alpha$ line in both
polarisation and polarisation angle spectra. The line effect is as
broad as the emission line and it appears as a linear excursion in the
{(\it Q, U)} diagram, the contribution of interstellar polarization
has likely introduced a signature in the position angle, but will have
only added a constant value in the {\it (Q,U)} diagram, which still
shows a straight line. An intrinsic polarisation line effect is seen
across the H$\alpha$ line in HD 163296, the effect is narrower than
depolarisation line effect and there is a flip in polarisation across
the line. This flip appears as a loop when it is mapped on the {(\it
  Q, U)} diagram. MWC 863 shows two different line effects, a McLean
line effect is seen across the absorptive component while the emission
line displays a narrow polarization which is identified with intrinsic
polarisation (in the following sections, the classificaiton criteria
are outlined). Fig. \ref{karim} also shows the intrinsic polarisation
angle which is measured from the slope of the line effect in the {(\it
  Q, U)} diagram. For the depolarisation line effect, the angle is
measured from the (unpolarized) line to the (polarized) continuum
while for the intrinsic polarisation and McLean effect it is measured
from continuum to (polarized) line (see the discussion in
\citealt{ababakr16}). Although the direction for both the
depolarization (HD 240010) and intrinsic polarization (HD 163296) are
the same in the respective {\it QU} graphs, the intrinsic polarization
angles differ, as this angle is measured from the line center to the
continuum in the case of the depolarization, it is measured in the
opposite direction for the intrinsic polarization.

\subsection{Statistical Results}

We present here H$\alpha$ spectropolarimetric results from a
sample of 56 HAeBe stars combined from this work and the
literature. We begin with discussing the observed line effect and its
type, before we focus on the magnitude and width of the line effect.

The H$\alpha$ spectropolarimetric results of each target are listed in
Table \ref{sss}. Columns 3 \& 6 list the spectroscopic
characterization of Stokes I, the intensity spectrum.  The line
polarimetric properties of each target are tabulated in columns 7-12
\& 15. Finally, the continuum spectropolarimetric measurements are
listed in columns 13 \& 14. As can be seen in Fig \ref{line_effect}
and Table \ref{sss}, a line effect is detected across H$\alpha$ in
42/56 objects (75 $\pm$ 6\%, errors reflecting the 1$\sigma$
confidence interval of a sample proportion), divided equally between
21 HBe and 21 HAe stars. 14 objects (25 $\pm$ 6\%) do not show any
signs of a line effect.
The detection rate for Herbig stars of
spectral type B0-B7 is 66 $\pm$ 9\% while that of the later type
(i.e. all other) objects is 85 $\pm$ 7\%, a difference that is close
to 3$\sigma$.

\begin{figure*}
        \centering
       \includegraphics[width=\textwidth]{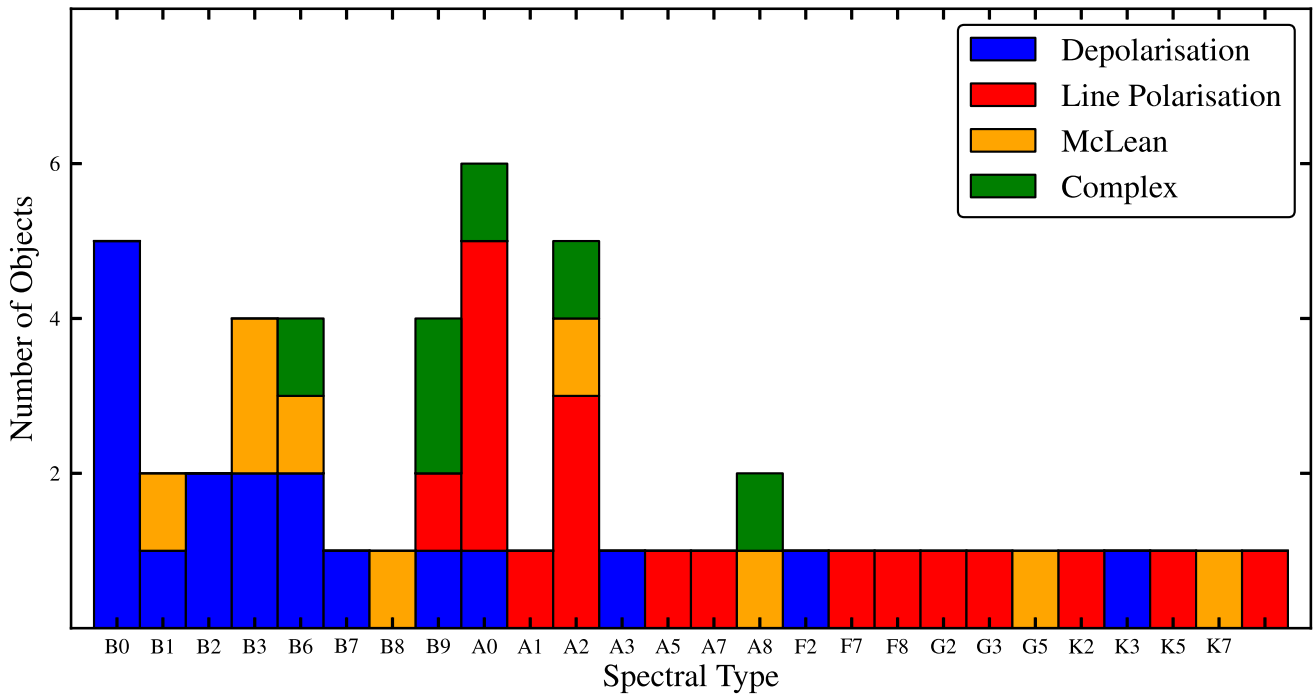}
        
   \caption[Type of line effects versus spectral type]{The figure
     represents the type of the observed line effect across H$\alpha$
     line as a function of spectral type of a sample of the 51 HAeBe
     and T Tauri objects that exhibit a line effect across the
     polarization. Note that the B4 and B5 spectral type bins which
     are in the previous figure are now unpopulated, as their (only)
     members do not show a line-effect.}
     \label{line_type_tauri}
\end{figure*}

\subsubsection{Line effect signatures}

The decision on whether to classify a line effect as depolarization or
intrinsic polarization can sometimes be subjective.  To avoid such
biases, we aim to differentiate between the depolarisation and the
line polarisation effect in a quantitative manner. To this end, the
width of the line effect can be used as proxy (see Table
\ref{sss}). The method was first used by \citet{Vink2002, vink05a} to
categorize the line effects. They statistically classified stars
according to the fractional width
[$\Delta\lambda(pol)/\Delta\lambda(I)$] at which the polarisation
changes across the line. This quantity measures the width of the
polarization over the line divided by the width of the line
itself. Generally a wide fractional width is associated with
depolarisation line effect, whereas the width of line polarisation
effect is often narrower than the depolarisation. Both
$\Delta\lambda(I)$ and $\Delta\lambda(pol)$ are measured at Full Width
at Zero Intensity (FWZI).  If the line effect is detected across the
absorption component of the emission line then it is considered as a
McLean line effect. The line effect across the emission line is
classified based on two criteria; the value of the fractional width
and whether there is a flip in the polarisation and PA spectra or
not. If the fractional width is equal to or larger than 0.7 and there
is no flip in the polarisation and PA spectra across the emission line
then the line effect is consistent with depolarisation. On the other
hand, if the fractional width is equal to or larger than 0.7 and a
flip is observed in either the polarisation or PA spectra then the
line effect is considered to be due to intrinsic polarisation. In
addition, if the fractional width is smaller than 0.7 then the line
effect is also consistent with polarisation. In some cases two
different line effects, a McLean line effect across the absorption and
either depolarisation or intrinsic polarisation across the emission
line, can be observed for the same object.

As reported before and mentioned in the Introduction, a depolarisation
is more common in HBe stars, in particular in the early HBe stars,
while in HAe stars intrinsic line polarisation is the dominant line
effect. To see at what spectral type the line effect switches from
line polarisation to depolarisation, we add 9 T Tauri stars to the
sample from \citet{vink05a} and the result is shown in
Fig. \ref{line_type_tauri}. The figure demonstrates that the intrinsic
line polarisation is the dominant effect in T Tauri and late HAe stars
while most early HBe stars show a depolarisation. There is also a
number of HBe stars with a McLean line effect. Based on the large fraction
of line effects at spectral types earlier than B7 and its absence
beyond that, it would appear that the line effect changes from
intrinsic line polarisation to depolarisation around B7-B8 spectral
type.

\subsubsection{Line effect magnitude}

\begin{figure}
        \centering
       \includegraphics[width=7.5cm]{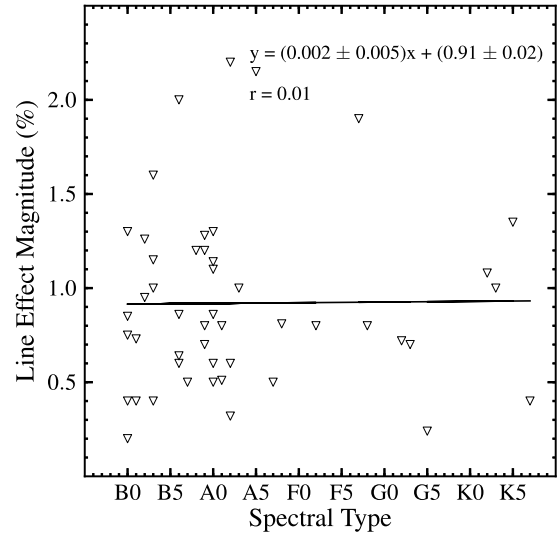}        
       \caption[Magnitude of line effect versus spectral type]{The
         figure shows the correlation between the magnitude of the
         observed line effect and the spectral type. The black line
         denotes the best-fitting straight line to the data. The best
         fit values are shown in the inset, as well as the correlation
         coefficient. The error on the magnitude of the line effect is
         below 10\%.}
            \label{magni}
\end{figure}

\begin{figure}
        \centering
       \includegraphics[width=7.5cm]{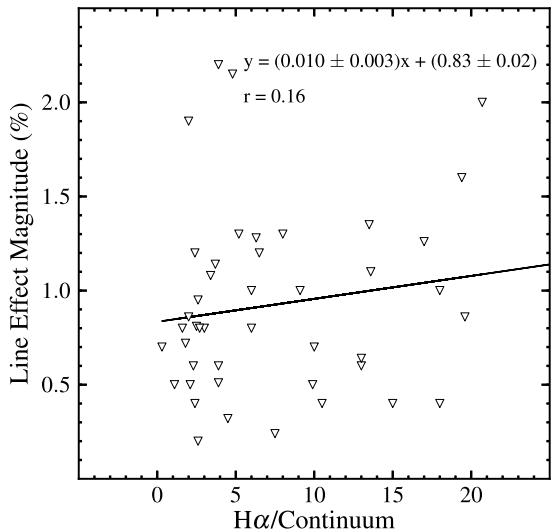}
        
   \caption[]{The correlation between the magnitude of the line effect
     and the strength of H$\alpha$. The black line denote the
     best-fitting line to the data. The error on the magnitude of the line
     effect is below 10\%, and the error of the peak of H$\alpha$ to
     the continuum is less than 5\%.}
   \label{ls_ew}
\end{figure}

Bearing in mind that the typical magnitude of the continuum
polarization, and thus line-effect, caused by electron scattering is
expected to be 1-2\% \citep{cassinelli1987}, we investigated the
magnitude of the line effect for all the objects that show a clear
line effect.  We measured the strength of the line effects directly from the
  {(\it Q, U)} diagram. It was taken as the distance between the continuum
  polarization, which is visible by a cluster of points,
  and the line centre.  The error was estimated to be typically 10\%.
The results are tabulated in Table \ref{sss} and are also shown in
Fig. \ref{magni}. The polarization ranges from $\sim$0.3$\%$ to
$\sim$2.0$\%$ with an average of $\sim$0.9 $\%$. Only R Mon shows a
magnitude of $\sim$10$\%$ which is not expected from electron
scattering close to the star and is due to observational effects as
the object is spatially resolved in these observations (see the
discussion in \citealt{ababakr16}). We have therefore discarded R Mon
from the final results.  As shown in Fig. \ref{magni}, the strength of
the line effect does not show any correlation with spectral type. To
investigate whether the strength of the emission lines is correlated
with the magnitude of the line effect, we plotted the magnitude of the
latter as a function of the line peak to continuum of the
H$\alpha$ line (see Fig. \ref{ls_ew}). As can be seen in the figure
there is only a very weak correlation between them.

\subsubsection{Fractional width}

\begin{figure}
        \centering
       \includegraphics[width=7.5cm]{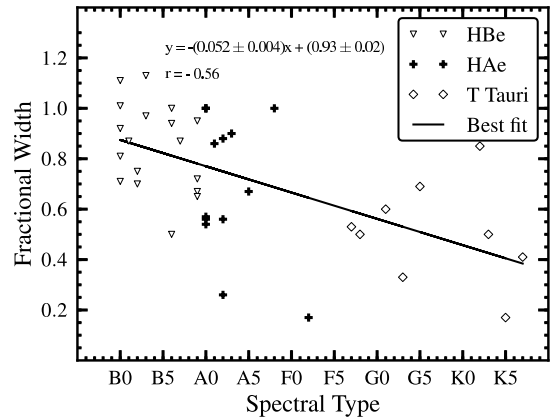}
        
   \caption[Fractional width of line effect]{The fractional width of
     the line effect ($\Delta\lambda(pol)/\Delta\lambda(I)$) is
     plotted against spectral type of a sample of HAeBe and T Tauri
     stars. The black solid line represents the best fit of the data,
     the correlation coefficient R=0.60, denoting a significant
     correlation, which is confirmed by the large, 10$\sigma$
     significance of the slope. The black solid circles with error
     bars represent the averages of early HBe, late HBe, early HAe,
     late HAe and T Tauri from early to late spectral type
     respectively.}

   \label{frac}
\end{figure}

\citet{Vink2002, vink05a} found that the fractional width
 tends to decrease towards late spectral type. We can now revise the
 relation by increasing the sample from 25 to 41 objects. The result
 is plotted in Fig. \ref{frac}, the figure shows that there is a
 significant correlation between the fractional width and the spectral
 type, with a correlation coeffient, r = $-$0.60. The slope of the
 best fit line between the fractional width against spectral type
 (counted as integers with B0=1, B1=2 etc) is determined at the
 10$\sigma$ level, providing another indication that the trend is
 real. The intrinsic scatter around the line does prevent us from
 making a conclusive statement whether there is a break in the
 relationship or whether it is continuous however. For example, when
 splitting the sample into early HBe, late HBe, early HAe and late
 type HAe stars, we find average values of 0.90$\pm$0.04,
 0.79$\pm$0.05, 0.74$\pm$0.07, 0.61$\pm$0.17 respectively (the errors
 are the scatter around the mean divided by the squareroot of the
 number of datapoints). The fractional widths measured for late HBe
 stars and early HAe stars are close and this might suggest they share
 a similar spectropolarimetric behaviour, but they themselves do not
 differ beyond the 2$\sigma$ level from either the early Be stars or
 the late HAe stars. What is clear, however, is that the trend from
 depolarization to intrinsic polarization with the spectral type is
 significant. While, in addition,  the early Herbig Be stars are distinctly
 different from the T Tauri stars for example.

\section{Discussion}

\subsection{Overall findings}

In the above we have investigated various observational aspects of the
linear spectropolarimetric properties of the young pre-Main sequence
Herbig Ae/Be stars.  This is the largest sample to have been studied
in this manner, and this allows us to confirm that there are distinct
differences between the lower mass Herbig Ae stars and the higher mass
Herbig Be stars. Indeed, we find evidence that the main distinction
occurs at the B7-B8 range. Furthermore, the Herbig Ae stars display a
similar behaviour as the solar mass T Tauri pre-Main Sequence
stars. The major statistical conclusions of this exercise can be
summarized as follows:

\begin{itemize}

\item The occurrence of the line effect in the entire sample is high,
  at 75\% (see e.g. Figure 3). When considering the sample of Herbig
  Ae and Herbig Be stars separately, we find that the occurrence in
  Herbig Be stars is smaller than in Herbig Ae stars.  This difference
  is amplified when splitting the sample in early B-type objects and
  the rest. The detection rate for Herbig stars of spectral type
  B0-B7 is 66 $\pm$ 9\% while that of the
  later type (i.e. all other) objects is 85 $\pm$ 7\%.  Hence, this
  difference is close to 3$\sigma$.

\item The appearance of the line effect is also different as a
  function of spectral type. Whereas the early type objects
  predominately display depolarization or McLean effects, the later
  type objects show intrinsic line polarization. This is very clear to
  see from Figure 4, where the break appears to occur around B7.
  Statistically, the overall change in the character of the line
  effect is also visible using the "fractional width" as a
  quantitative handle on the nature of the line effect. Figure 7 shows
  that Herbig Be stars have larger polarization widths than later type
  objects. The trend is significant with a slope at the 10$\sigma$
  level.

\item The strength of the line effect (in terms of per cent
  polarization) is of order 1 \% and is independent of spectral
  type. However, there appears to be a weak trend in that stronger
  H$\alpha$ emission lines have a slightly larger effect.

\end{itemize}

The main result of the present study is that the difference between
Herbig Ae and Herbig Be stars is now more robust, not only the
detection rates are, statistically, shown to be different, but this
3$\sigma$ effect is underpinned by the fact that the nature of the
line effects is different as well. In the following, we discuss what
these findings mean for the formation mechanism of low, intermediate
and high mass stars, and why the line effect is equally strong in these
classes of object.

\subsection{The origin and nature of the line effect}

As discussed earlier, the intrinsic line polarization observed towards
the cooler objects can be explained by compact emission such as
accretion shocks on the stellar surface scattering off circumstellar
material, which is found to be consistent with a disk.  On the other
hand, the line depolarization and McLean effect can be best explained
with the presence of a small circumstellar disk. As shown in the case
of classical Be stars, most of the polarization originates from a
region within a few stellar radii, where the electron densities are
highest. It may be surprising then that there is only a weak
  correlation between the magnitude of the line effect and the
  strength of the H$\alpha$ line (or spectral type), as one could
  expect a stronger emission line to be associated with more
  ionization and more free electrons and thus a larger polarization
  and larger line effect.

This suggests that the detection of the line effect only depends on
the geometry of the scattering agents and the geometry of the line
emitting region, whereas, in the optically thin limit, the
polarization itself depends on the density of the scatterers.  For the
depolarisation line effect, the free electrons in the ionised region
around the stars polarise the continuum photons while the emission
photons are unpolarised.  In this context, it is useful to note
  that for example \citet[e.g. their Figure 1]{Mottram07} detected a
  clear line effect across the H$\beta$ emission of Herbig Ae/Be
  objects. For some stars the lines are so faint that the emission
  does not even reach the photospheric continuum. However, in such
  cases, the line can still be several times stronger than the
  underlying photospheric emission. This is because the photospheric
  absorption lines' minimum can be as low as 0.2 times the continuum
  level, and when the emission reaches the continuum level, it will
  have first filled up the underlying absoprtion. Most of the observed
  emission at these wavelengths will then be the unpolarized line
  emission. The maximum observable line effect, being the difference
  between continuum and line emmision, is therefore reached already
  for weak lines and no, or hardly any, further changes in polarization
  is observed for progressively stronger lines.

In the case of the intrinsic line polarisation in our objects, the
line effect is due to the fact that compact line emission (such as
from individual accretion hot-spots or accretion funnels) scatters off
circumstellar material.  The cause for the ionization leading to the
line emission and that responsible for the free electrons in the disk
may be linked, but emission and polarization do not necessarily have
to be correlated. For example the line emission arises from localized
accretion hot spots and funnels in the magnetospheric accretion
paradigm, while the circumstellar disk itself would be generally
ionized due to the stellar photosphere and accretion luminosity.
Indeed, it has been pointed out it is not yet settled whether in these
situations the disk scattering material are free electrons, neutral
hydrogen or dust (see e.g. \citealt{woodbrown1994, vink2015} for
discussions).  In any case, no trend of the strength of the line
effect with line strength itself needs to be expected.  In conclusion,
in the above situations we can understand that the line effect
strength is independent of the line strength itself. For the McLean
effect, which we observe in a number of objects, this may not
necessarily be the case.  The number of objects with the McLean effect
is rather small, 8 objects, but there is no correlation between the
magnitude of the line effect and the strength of the emission line. As
the scattering is due to the inner regions, unrelated to the outflow
or infall itself, we suspect that this also leads to a line effect
independent of the line emission.

Finally, we address the question why the detection rate of the line
effect in the Herbig Be stars is lower than for the Herbig Ae stars;
in the case of line depolarization, the effect is a strong function of
inclination of the system, A larger disk inclination results in a
stronger line effect \citep{wood1993}. When the disk is pole-on for
example, the system is circular on the sky and all polarization
vectors will cancel out, resulting in a net zero polarization and no
difference in polarization between line and continuum.  A 100\%
detection rate will therefore not be expected at all for a sample of
objects distributed at random inclinations, and the current detection
rate is similar to those of classical Be stars (see
e.g. \citealt{oudmaijer07}).  A difference in the case of intrinsic
polarization is that polarization can be seen even at low, face-on,
inclinations of the disk. This is mainly due to the fact that the
compact emission is anisotropic (be it due to accretion hot spots or
funnels). As a consequence only part of the disk will be
illuminated. This results in a net observable polarization, and in
passing we note that this also explains the higher fraction of
line-effects for later type stars which predominately exhibit the
intrinsic polarization line-effect.

\subsection{On the formation of intermediate and massive stars}

Our findings indicate that a break between the spectropolarimetric,
and possibly accretion, properties of high and low mass objects occurs
around the B7-B8 spectral type. This is found in both the detection
statistics and, especially, the nature of the line effect. The early Herbig Be
stars have a lower detection rate and show predominately the
depolarization and McLean effects, indicative of circumstellar disks.

The detections are more numerous for the later B-type and A-type
objects. In addition, as can be seen in Fig.\ref{line_type_tauri}, T
Tauri stars and, especially late, HAe stars share the same
H$\alpha$ intrinsic polarisation line effect. This effect can be
explained by scattering photons originating from a compact source,
where the accretion take place on to the star, off the circumstellar
electrons. The similarities in the spectropolarimetric properties
between T Tauri stars - which are known to undergo magnetospheric
accretion - and the late-type Herbig Be and Herbig Ae stars suggest
this mechanism also acts on these intermediate mass stars
(cf. \citealt{Vink2002} and references in the Introduction).

A complication is whether HAe stars have sufficient magnetic fields to
facilitate MA or not. \citet{wade05, wade07, alecian13, hubrig09,
  hubrig13} detected magnetic fields ($\sim$ a few hundred G) in a few
HAeBe stars, mostly HAe and late HBe stars, but it is as of yet not
clear whether this would be sufficient to drive
accretion.

Currently, there is no well-explored theory that explains the
accretion of material onto the highest mass Young Stellar Objects
however. Even the most recent, sophisticated star formation models are
not able to simulate the fine detail required to probe the accretion
process from parsec scales via an accretion disk to the stellar
surface.  For example, \citet{rosen16} explicitly mention that the
material is not followed in the inner 80 au.  Given that our
spectropolarimetric evidence points towards circumstellar disks that
are present at very small scales, the logical, direct conclusion we
can draw is that the disk is not truncated, but reaches all the way
down to the stellar surface.  In this situation, the so-called
Boundary Layer accretion is a viable mechanism to explain the growth
of massive stars. The BL is a thin annulus close to the star in which
the material reduces its (Keplerian) velocity to the slow rotation of
the star when it reaches the stellar surface, and it is here that
kinetic energy will be dissipated. It has been explored for Herbig
Ae/Be stars by \citet{blondel06}, and the BL mechanism has also been
explicitly suggested a number of times to act in Herbig Be stars
(e.g. \citealt{mendigutia11, cauley2015, hartmann2016,beltran2016}),
however, details are yet to be worked out for the more massive Herbig
Be stars.

\section{Conclusions}

This work presents the spectropolarimetric results of a sample of 56
HAeBe stars which is the largest linear spectropolarimetric sample of
HAeBe stars that has been published to date. The main findings are as
follows:

%\begin{itemize}
\begin{enumerate}[(i)]

\item Most HAeBe stars show a sign of line effect which is interpreted
  by the presence of a circumstellar disk. The detection rate of the
  line effect is 75 $\%$ (42/56) in the sample of 56 objects that have
  been observed spectropolarimetrically. 

\item The magnitude of the line effect is of order of 0.3-2
  $\%$. There is no correlation between this magnitude and spectral
  type. A very weak correlation is seen between the magnitude of the
  line effect and the strength of H$\alpha$ line. We can explain this
  both in terms of the line depolarization and intrinsic
  polarization. The detection of the line effect does not rely on the
  strength of the emission line but on the geometry of circumstellar
  environment.

\item The Herbig Be type stars have
  a significantly lower detection rate than the Herbig Ae stars, with
  a break around spectral type B7-B8.

\item Most of the HBe stars' signatures are consistent with a
  depolarisation line effect. In contrast, intrinsic line polarisation
  is more common in HAe and T Tauri stars. It seems late HBe and early
  HAe stars are at the interface between the two line effects, also
  indicating a break in spectropolarimetric properties around
  B7-B8. The similarity between T Tauri, HAe stars and late Herbig Be
  stars suggest that the latter are forming via magnetospheric
  accretion.

\end{enumerate}
%\end{itemize}

The interface between HBe and HAe is possibly where the accretion
mechanism switches from magnetospheric accretion to another
process. Given the fact that the Herbig Be stars are non-magnetic and
surrounded by small scale disks, it will be interesting to consider
and work out in details the Boundary Layer model as a means for the
continued accretion and growth of more massive stars.

\section*{Acknowledgements}

The allocation of time on the William Herschel Telescope was awarded by PATT, the United Kingdom allocation panel. KMA was supported by the
UK Science and Technology Facilities Council (STFC). This research has made use of the SIMBAD data base, operated at
CDS, Strasbourg, France.

%%%%%%%%%%%%%%%%%%%%%%%%%%%%%%%%%%%%%%%%%%%%%%%%%%



\bibliographystyle{mn2e}

%\begin{thebibliography}{99}
%\end{thebibliography}
\bibliography{references.bib}
%\bibitem[Vink et al.(2005)]{2005A&A...430..213V} Vink, J.~S., Harries, T.~J., \& Drew, J.~E.\ 2005, \aap, 430, 213 
%%%%%%%%%%%%%%%%%%%%%%%%%%%%%%%%%%%%%%%%%%%%%%%%%%

%%%%%%%%%%%%%%%%% APPENDICES %%%%%%%%%%%%%%%%%%%%%

\begin{landscape}
\begin{table}
 %\centering
 %\begin{minipage}{140mm}
\caption[Log of H$\alpha$ line polarimetry of the FORS2 and ISIS observations]{The H$\alpha$ spectropolarimetric results of the largest sample of HAeBe stars. The measurements are taken in the R band. Columns 5 \& 6 list the Stokes (I) characteristics EW and line to continuum ratio, the error in both are typically less than 5\%. Columns 7-12 list line spectropolarimetric characteristics of each targets. Column 7 shows whether there is a line effect across the H$\alpha$ line or not and the magnitude of the line effect is presented in column 8. The typical error is 10\%. $\Delta\lambda(pol)$ and the fractional width $\Delta\lambda(pol)$/$\Delta\lambda(I)$ are listed in column 9 \& 10, $\Delta\lambda(pol)$ and $\Delta\lambda(pol)$ are measured at full width zero intensity (FWZI). $\Delta\lambda(pol)$ is the width over which the polarisation changes across the line, column 11 lists whether there is a flip in polarisation or polarisation angle or not across the line. The classification of the line effect is presented in column 12. The continuum polarisation and polarisation angle are listed in columns 13 \& 14 and the intrinsic polarisation angle is listed in column 15.}

\begin{tabular}{l l l l r r  r r  r r l  l r r r l }
\hline

  No.&Object &Obs date& Spec.  &H$\alpha$~EW  &  Line/cont. &Line &Mag.(\%)& $\Delta\lambda(pol)$& \underline{$\Delta\lambda(pol)$} &Flip &  Class.&$P_{cont}(\%)$&$\theta_{cont}^{\circ}$& $\theta_{intr}^{\circ}$&Ref. \\
 && & type    & \rm\AA& &effect& &  \rm\AA &  $\Delta\lambda(I)$ && &&  \\
\hline 
1&HD 163296     &04-08-15 &    A1  &-16.4&   3.9  & Y & 0.51 &15.0  &0.86&Y&P&0.162~$\pm$~0.005&48.88~$\pm$~0.90& 144$\pm$10&1   \\ %herbig cata both
&&01-04-12 & &-14.9&  3.8  &Y& &19.5&  1.05  & Y  &   P &0.70~$\pm$~0.01&1.60~$\pm$~0.20&112$\pm$10& 2       \\ 
2&MWC 610     & 04-08-15&    B2     &2.0& 1.01 & N  & -  & -&-&-&-&2.507~$\pm$~0.006&81.86~$\pm$~0.07&-& 1        \\%s cauley m simbad
3&MWC 342      & 04-08-15 &    B(e)  &-264.5&  54.0   & Y  & 0.75  &17.5&0.71&N&D &0.736~$\pm$~0.016&102.17~$\pm$~0.57 &180$\pm$5&1\\ %simbad both delta intensity %2 double fip caused by absorption
4&HD 200775    & 04-08-15 &    B3   &-75.0&  10.5  & Y& 0.40  &29.0& 1.10&N&D/M&0.782~$\pm$~0.005&93.46~$\pm$~0.19& 88$\pm$5&1    \\ %s herbig m simbad double fip caused by absorption
&&18-12-99& &-63.0&8.5&Y&0.50&45.0&1.13&N&D/M&0.81$\pm$0.01&96$\pm$1&93$\pm$2&3,6\\ %vink02 error in angle less than 1 degree
&&13-12-03& &-47.0&8.2&Y&0.35&-&-&N&D/M&0.801$\pm$0.003&93.6$\pm$0.1&90&4\\ %vink05
5&HD 203024    & 04-08-15 &    A1    &7.6& 0.6  &N &  -  &- &-&-& -&0.496~$\pm$~0.007&49.23~$\pm$~0.42& - & 1  \\ %herbig both
6&V361 Cep     &04-08-15 &    B3   &-32.5&  5.9  &  N&- & -&-&-&-&0.846~$\pm$~0.007&87.84~$\pm$~0.25&-&1 \\%s cauley m simbad
7&HD 240010    &04-08-15&    B1   &-14.3&  2.4  &  Y & 0.40 &27.0& 0.87&N&D&3.018~$\pm$~0.008&73.12~$\pm$~0.08&100$\pm$5 &1  \\ %simbad both
8&V374 Cep     & 04-08-15  &    B0  &-33.6& 5.2    & Y & 1.30&27.5&0.81&N&D &4.841~$\pm$~0.008&128.46~$\pm$~0.05&169$\pm$5&1    \\%herbig both
9&MWC 863     & 05-08-15&    A2    &-1.6&  3.0 & Y &0.80 &15.0&1.00&N&M/P &5.319~$\pm$~0.004& 57.25~$\pm$~0.02& 55$\pm$5& 1  \\ %herbig both
10&V718 Sco    & 05-08-15  &   A7   &6.3&  1.1  &Y & 0.50&12.0& 0.72&N&M &0.457~$\pm$~0.007&83.07~$\pm$~0.44& -&1 \\%s cailey m simbad
11&HD 141569   & 05-08-15   &    A0  &4.2& 1.3    & N & -  & -&-&-& -&0.639~$\pm$~0.004 &86.91~$\pm$~0.18&- & 1   \\ %s herbig m simbad
&   & 27-12-01   &      &-5.5& 1.4    & N & -  & -&-&-& -&0.647~$\pm$~0.005 &85.2~$\pm$~0.2&- & 1   \\ %vink05
12&MWC 300     &05-08-15 &    B1  &-128.0&  71.0   & Y &0.73 &3.5& 0.29&N&M&4.467~$\pm$~0.016&57.58~$\pm$~0.10 &-&1\\% s herbig m simbad
13&MWC 953     & 05-08-15 &    B2   &-34.1&  12.8  & N&  - &- &-&-&-&2.555~$\pm$~0.008&75.60~$\pm$~0.10&- &1 \\ %simbad both
14&MWC 623      &05-08-15   &    B4(e)  &-119.3&  38.0   &   N& - &- &-&-&-&2.548~$\pm$~0.011&53.50~$\pm$~0.12&-&1   \\ %simbad both
15&V1977 Cyg   &05-08-15 &    B8    &-31.2& 6.5  & Y &1.20&5.8& 0.34&N&M&2.774~$\pm$~0.008&168.89~$\pm$~0.08 &-& 1\\% s herbig m simbad
16&MWC 1080& 18-12-99& B0&-101.0&18.0&Y&0.40&50.0&1.11&N&D&1.73$\pm$0.01&77$\pm$1&&3\\ %vink02
&    &05-08-15   &     &-126.0&  19.0   &  Y &0.30 &7.8&0.29&N&M&1.601~$\pm$~0.007& 82.24~$\pm$~0.12 &-&1  \\ %s herbig m simbad
&    &29-09-04   &     &-100.0&   30.0  &  Y&0.30 &-&-&N&D&1.51~$\pm$~0.01& 77.2~$\pm$~0.1 &-&6  \\ %s herbig m simbad %
&    &09-11-08   &     &-&  - &  -&- &-&-&-&-&1.7~$\pm$~0.1& 78~$\pm$~1 &-&7  \\%wheelwright
17&HD 144432    &05-08-15   &    A8   &-2.5& 2.5   &   Y & 0.81 &12.5&1.00&N&D/M&0.432~$\pm$~0.004&12.92~$\pm$~0.25&175$\pm$10&1   \\% s herbig m simbad may be flip

%%%%
18&PDS 27 &04-02-12 &   B3 &-120.8 & 18.0&Y&1.00&   18.0   &    0.57       &N &M &8.80~$\pm$~0.01 & 18.40~$\pm$~0.04& 77$\pm$10&2         \\
19&PDS 37  &05-02-12 &    B3 &-122.6 &  19.4 &Y&1.60&   15.5   &     0.55    &N&  M& 5.20~$\pm$~0.01&130.00~$\pm$~0.06&56$\pm$10& 2      \\
20&PDS 133& 21-01-12&  B6 & -94.5&20.7 &Y&2.00& 6.0 &   0.28 & N  & M & 2.30~$\pm$~0.04 & 37.70~$\pm$~0.50&126$\pm$10&2        \\
21&HD 98922& 16-01-12&  B9&-17.9&  6.0 &  Y&0.80 &9.8 &   0.72 & N&  D/M & 0.40~$\pm$~0.01 &174.90~$\pm$~0.30& 22$\pm$10&2           \\ % no flip
22&GU CMa & 07-01-12&  B2 &-11.1& 2.6 &Y &0.95& 12.8& 0.70 &  N & D &1.60~$\pm$~0.01  &14.60~$\pm$~0.01&127$\pm$10&2       \\
&& 11-01-95&  &-14.0& 3.0 &N &- &- & - & -  & - &1.15~$\pm$~0.01  &19~$\pm$~1&-&5       \\
&& 08-11-08&  &-& 2.3 &Y & 0.30&- & - & N  & D &0.8~$\pm$~0.1  &24~$\pm$~1&-&7       \\%wheelwright
&& 10-12-03&  &-9.3& 2.2 &N & -&- & - & -  & - &1.726~$\pm$~0.006  &27.0~$\pm$~0.1&-&4       \\%vin05
23&CPD-485215&20-01-12&   B6& -100.8& 19.6 &Y&0.86&17.0& 1.00&  N &  D&1.70~$\pm$~0.01& 9.50~$\pm$~0.20&62$\pm$10&2      \\
24&HD 85567 &30-12-11 &  B7& -43.5& 9.9&Y& 0.50&17.0&     0.87    &N& D &0.38$\pm$0.01&133.90$\pm$0.60&31$\pm$10&2      \\
&&30-03-12&&-42.1&10.2&Y?& - &-&   -    &- & -&0.26$\pm$0.02&134.50$\pm$2.10&- &2      \\ %less than 3 sigma ignore it
25&V380 Ori&03-01-12&A0&-71.1& 13.6&Y&1.10 &7.0&  0.54   &N& P&0.70$\pm$0.01&98.40$\pm$0.06&10$\pm$10&2   \\
&&12-10-11&   &-82.7&15.5 &N&- & - & - &  -   &- &0.70$\pm$0.01&95.15$\pm$1.00&-& 2     \\
&&31-12-96&&-79.0& 14.0&N& -&-&  -  &-& -&1.26$\pm$0.01&96$\pm$1&-&5\\ %oudmaijer
26&HD 104237  &16-01-12 & A7 & -27.2 &6.5 &Y?& -  &  -  &    -    & -& Complex  &0.35~$\pm$~0.01 &174.60~$\pm$~0.60&-& 2     \\ %less than 3 sigma ignore it
27&BF Ori  &12-10-11&   A2 & 1.0  & 1.9&Y? &  - & -&   -     & - &Complex& 0.70~$\pm$~0.01& 44.00~$\pm$~0.06&-&2 \\ %ignore equal 3 sigma
&  &09-11-08&    & -  &1.8&N &  - & -&   -     & - &-& 0.6~$\pm$~0.1& 58~$\pm$~1&-&7 \\ %wheelwright
28&MWC 137& 20-12-99& B0&-404.0&70.0&Y&0.85&55.0&0.92&N&D&6.07$\pm$0.01&160$\pm$1&25$\pm$10&3\\ %vink02
&& 31-12-96& &-550.0&83.0&Y&0.69&28.6&1.06&N&D&6.11$\pm$0.01&162$\pm$1&30$\pm$5&5\\ %oudmaijer
29&BD+40 4124 &19-12-99& B2&-113.0&17.0&Y&1.26&30.0&0.75&N&D&1.21$\pm$0.01&8$\pm$1&83$\pm$8&3\\ %vink02
\hline
\label{sss}
\end{tabular}
%\begin{minipage}{\columnwidth}
%D: Depolarisation, P: Polarisation, M: McLean
%\end{minipage}
%\end{minipage}
\end{table}
\end{landscape}

\begin{landscape}
\begin{table}
\setcounter{table}{1} 
 %\centering
 %\begin{minipage}{140mm}
\caption[Log of H$\alpha$ line polarimetry of the FORS2 and ISIS observations]{continued.}

\begin{tabular}{l l l l r r  r r  r r l  l r r r l}
\hline

  No.&Object &Obs date& Spec.  &H$\alpha$~EW  &  Line/cont. &Line &Mag.(\%) & $\Delta\lambda(pol)$& \underline{$\Delta\lambda(pol)$}  &Flip&  Class.&$P_{cont}(\%)$&$\theta_{cont}^{\circ}$& $\theta_{intr}^{\circ}$&Ref. \\
 && & type    & \rm\AA& &effect& &  \rm\AA &  $\Delta\lambda(I)$ && &&  \\
\hline
 
30&MWC 147 &18-12-99& B6&-60.0&13.0&Y&0.60&18.0&0.50&-&Complex&1.05$\pm$0.01&100$\pm$1&-&3\\ %vink02
&&30-12-96& &-63.0&11.0&Y&0.25&6.8&0.33&N&-&1.06$\pm$0.01&102$\pm$1&-&5\\ %oudmaijer check text for intrinsic angle
&&09-11-08& &-& -&Y&-&-&-&N&-&1.0$\pm$0.1&100$\pm$1&-&7\\ %wheelwright
31&HD 58647 &19-12-99& B9&-8.6&2.4&Y&1.20&10.0&0.67&N&Complex&0.14$\pm$0.01&127$\pm$1&-&3\\ %vink02
32&MWC 120 &19-12-99& B9&-29.0&6.3&Y&1.28&13.0&0.65&N&P&0.35$\pm$0.01&76$\pm$1&-&3\\ %vink02
&&11-01-95& &-17.0&4.9&Y&-&-&-&N&-&0.29$\pm$0.01&121$\pm$1&-&5\\ %oudmaijer
&&29-09-04& &-20.0&-&Y&-&-&-&&-&0.40$\pm$0.01&115.4$\pm$0.2&&6\\ %mottram
33&MWC 789 &19-12-99& B9&-46.0&10.0&Y&0.70&21.0&0.95&Y&Complex&0.92$\pm$0.01&174$\pm$1&178$\pm$10&3\\ %vink02 how intrinsic angle
34&AB Aur & 18-12-99&A0&-40.0&8.0&Y&1.30&17.0&1.00&N&Complex&0.11$\pm$0.01&54$\pm$1&160$\pm$5&3\\ %vink02 how intrinsic angle
35&XY Per &20-12-99& A2&-6.7&2.3&Y&0.60&15.0&0.88&Y&P&1.60$\pm$0.01&132$\pm$1&-&3\\ %vink02
& &10-11-08& &-&-&N&-&-&-&-&-&1.4$\pm$0.1&128$\pm$1&-&7\\ %wheelwright
36&MWC 480 &19-12-99& A5&-21.0&4.8&Y&2.15&12.0&0.67&Y&P&0.38$\pm$0.01&52$\pm$1&-&3\\ %vink02
&&30-09-04& &-11.0&-&Y&-&-&-&-&P&0.20$\pm$0.01&64.9$\pm$1.3&-&6\\ %mottram
&&26-12-01& &-20.5&4.9&Y&-&-&-&-&M&0.176$\pm$0.005&63.0$\pm$0.8&55&4\\ %vink05 new check angle
37&HD 244604 &19-12-99& A2&-14.0&4.5&Y&0.32&10.0&0.56&Y&P&0.44$\pm$0.01&119$\pm$1&-&3\\ %vink02
38&T Ori &20-12-99& A2&-12.0&3.9&Y&2.20&5.0&0.26&N&P&0.39$\pm$0.01&97$\pm$1&-&3\\ %vink02
39&HD 245185 &20-12-99& A0&-17.0&3.7&Y&1.14&10.0&0.56&Y&P&0.2$\pm$0.011&168$\pm$1&-&3\\ %vink02
40&IL Cep&08-11-08& B3&-20.0&3.5&N&-&-&-&-&-&4.24$\pm$0.01&102$\pm$1&-&3\\ %vink02
&&19-12-99& &-&-&N&-&-&-&-&-&4.3$\pm$0.01&100$\pm$1&-&7\\ %wheelwright may be no line effect
41&MWC 758&19-12-99& A7&-17.0&4.0&N&-&-&-&-&-&0.07$\pm$0.01&179$\pm$1&-&3\\ %vink02
&&08-11-08& &-&-&N&-&-&-&-&-&0.5$\pm$0.1&47$\pm$1&-&7\\ %wheelwright
42&HD 35929&19-12-99& F2&-3.2&2.0&N&-&-&-&-&-&0.12$\pm$0.01&51$\pm$1&-&3\\ %vink02
43&SV Cep &12-12-03& A0&-11.0&2.0&Y&0.86&20.0&1.00&Y&P&1.293$\pm$0.011&67.6$\pm$0.2&-&4\\ %vink05
44&CQ Tau &20-12-99& F2&-2.7&1.6&Y&0.80&3.0&0.17&N&P&0.27$\pm$0.01&83$\pm$1&-&3\\ %not %vink05 it is vink02

45&MWC 158 &01-01-97& B6&-58.0&13.0&Y&0.64&15.75&0.94&N&D&0.65$\pm$0.01&154$\pm$1&155$\pm$5&5\\ %oudmaijer98
46&HD 87643&01-01-97& B3&-196.0&26.0&Y&1.15&31.0&0.97&N&D&0.75$\pm$0.01&164$\pm$1&-&5\\ %oudmaijer98 may be add intrinsic angle
&&31-12-96& &-186.0&25.0&Y&1.27&29.3&0.96&N&D&0.84$\pm$0.01&168$\pm$1&-&5\\ %oudmaijer98

47&W Ori&11-01-95& B3&-3.5&1.6&N&-&-&-&-&-&0.27$\pm$0.01&55.7$\pm$0.6&-&5,6\\ %mottram07 no line effect in h alphadata from %oudmaijer98
48&MWC 166&30-12-96& B0&-14.0&2.6&Y&0.20&24.4&1.01&N&D&0.49$\pm$0.01&44.9$\pm$0.1&136$\pm$4&5,6\\ %mottram07 data from %oudmaijer98
&&18-12-99& &-2.8&1.6&N&-&-&-&-&-&0.20$\pm$0.01&34$\pm$1&-&3\\
&&11-12-03& &0.59&0.75&N&-&-&-&-&-&0.164$\pm$0.003&36.8$\pm$0.4&-&4\\ %vink05 check it is in absorption unlike other observations

49&HD 179218&08-11-08 & A0&-&2.1&Y&0.50&8.0&0.57&N&P&0.5$\pm$0.1&111$\pm$1&$\sim$45&7\\ %wheelwright
50&HK Ori &08-11-08& A3&-&6.0&Y&1.00&15.0&0.90&N&D&1.4$\pm$0.1&113$\pm$1&169$\pm$6&7\\ %wheelwright
51&V586 Ori &09-11-08& A0&-&3.9&Y&0.60&13.0&1.00&N&D&0.9$\pm$0.1&84$\pm$1&81$\pm$8&7\\ %wheelwright

%nolineeffectoudmaijer
52&MWC 297&15-07-98& B1&-520.0&100.0&N&-&-&-&-&-&1.90$\pm$0.01&86$\pm$1&-&5\\ %oudmaijer98
53&HD 76534&11-01-95& B3&-7.5&2.0&N&-&-&-&-&-&0.52$\pm$0.01&124$\pm$1&-&5\\ %oudmaijer98
54&AS 116&01-01-97& B5&-90.0&18.0&N&-&-&-&-&-&1.41$\pm$0.01&30$\pm$1&-&5\\ %oudmaijer98
55&Hen 3-230&30-12-96& Be&-315.0&64.0&N&-&-&-&-&-&1.61$\pm$0.02&30$\pm$1&-&5\\ %oudmaijer98
56&HD 45677&11-01-95& B3&-200.0&35.0&Y&-&-&-&-&-&0.33$\pm$0.01&11$\pm$1&-&5\\ %oudmaijer98 check line effect
&&30-12-96& &-200&34&Y&-&-&-&-&-&0.14$\pm$0.02&143$\pm$3&-&5\\ %oudmaijer98
\hline
57&R Mon& 12-10-11&    B8 & -98.6 & 12.6 &Y &4.60 &30.5 &  1.00   & -&  -  &8.10~$\pm$~0.02&69.40~$\pm$~0.07&- &2        \\ % r mon discarded 
&&01-02-12&&-105.5 &12.9&Y &9.30 & 26.5& 0.82   &-& -& 12.15~$\pm$~0.01 &67.70~$\pm$~0.04&-&2\\
58&LKH$\alpha$ 218&31-12-96& B9&-20.0&6.0&N&-&-&-&-&-&1.91$\pm$0.02&19$\pm$1&-&5\\ %oudmaijer98 bad data
59&AS 477&19-12-99& A0&-19.0&4.4&N&-&-&-&-&-&0.43$\pm$0.01&56$\pm$1&-&3\\ %baddata
60&KMS 27&19-12-99& A0&-7.2&2.5&Y&-&-&-&-&-&0.13$\pm$0.01&52$\pm$1&-&3\\ %baddata
\hline
\label{sss}
\end{tabular}
\begin{minipage}{\columnwidth}
D: Depolarisation, P: Polarisation, M: McLean\\
References. 1: This work; 2:\citet{ababakr16}; 3: \citet{Vink2002}; 4:\citet{vink05a}; 5:\citet{1999MNRAS.305..166O}; 6: \citet{Mottram07}; 7: \citet{wheelwright11}\\
R Mon was discarded as the large magnitude of the line effect ($\sim$10\%) is not caused by electron scattering.\\
LKH$\alpha$, AS 477 and KMS 27 were discarded as they suffer from too low photon counts.\\ 
\end{minipage}
%\end{minipage}
\end{table}
\end{landscape}

\appendix
\section[]{Observed spectropolarimetric signatures}
Here we present the spectropolarimetric results across H$\alpha$ for the new observations.

\begin{figure*}
        
        \centering
       \includegraphics[width=16cm]{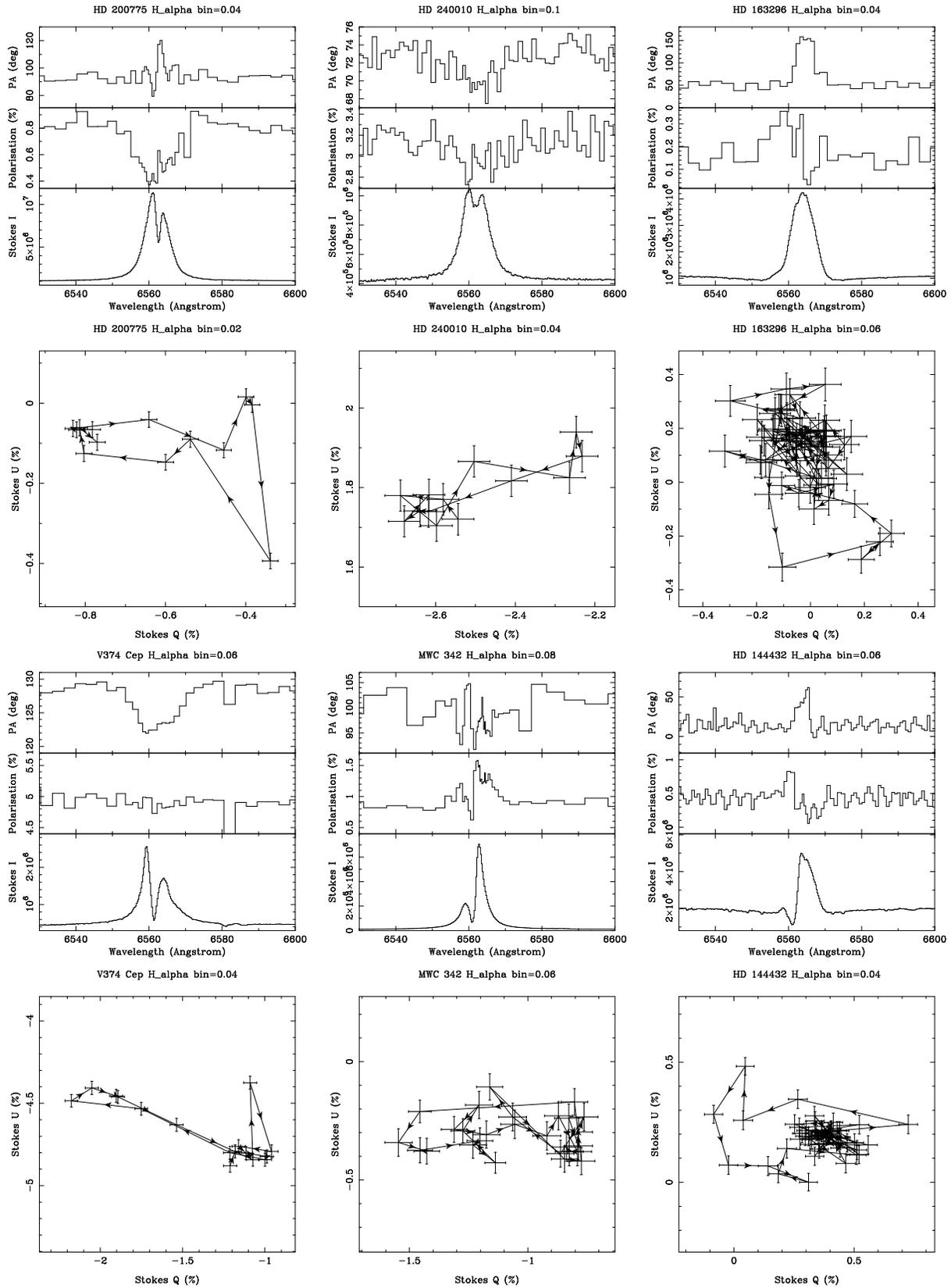}
   \caption{H$\alpha$ spectropolarimetry of the stars. The data are
     presented as a combination of triplots (top) and {(\it Q, U)} diagrams
     (bottom). In the triplot polarisation spectra, the Stokes
     intensity (I) is shown in the bottom panel, polarisation (\%) in
     the centre, while the position angle (PA) is shown in the upper
     panel. The Q and U Stokes parameters are plotted against each
     other below each triplot. The data are rebinned to a constant
     error in polarisation, which is indicated at the top of each
     plot. The arrows in the {(\it Q, U)} diagrams indicate the polarisation moves in and out of the line effect from blue to red wavelengths.}
    \label{halphafig}

\end{figure*}

  \begin{figure*}
\setcounter{figure}{0} 

\centering
        \includegraphics[width=16cm]{h2.ps}
           \caption{continued}
   \end{figure*}

 \begin{figure*}
\setcounter{figure}{0} 
        
        \centering
       \includegraphics[width=16cm]{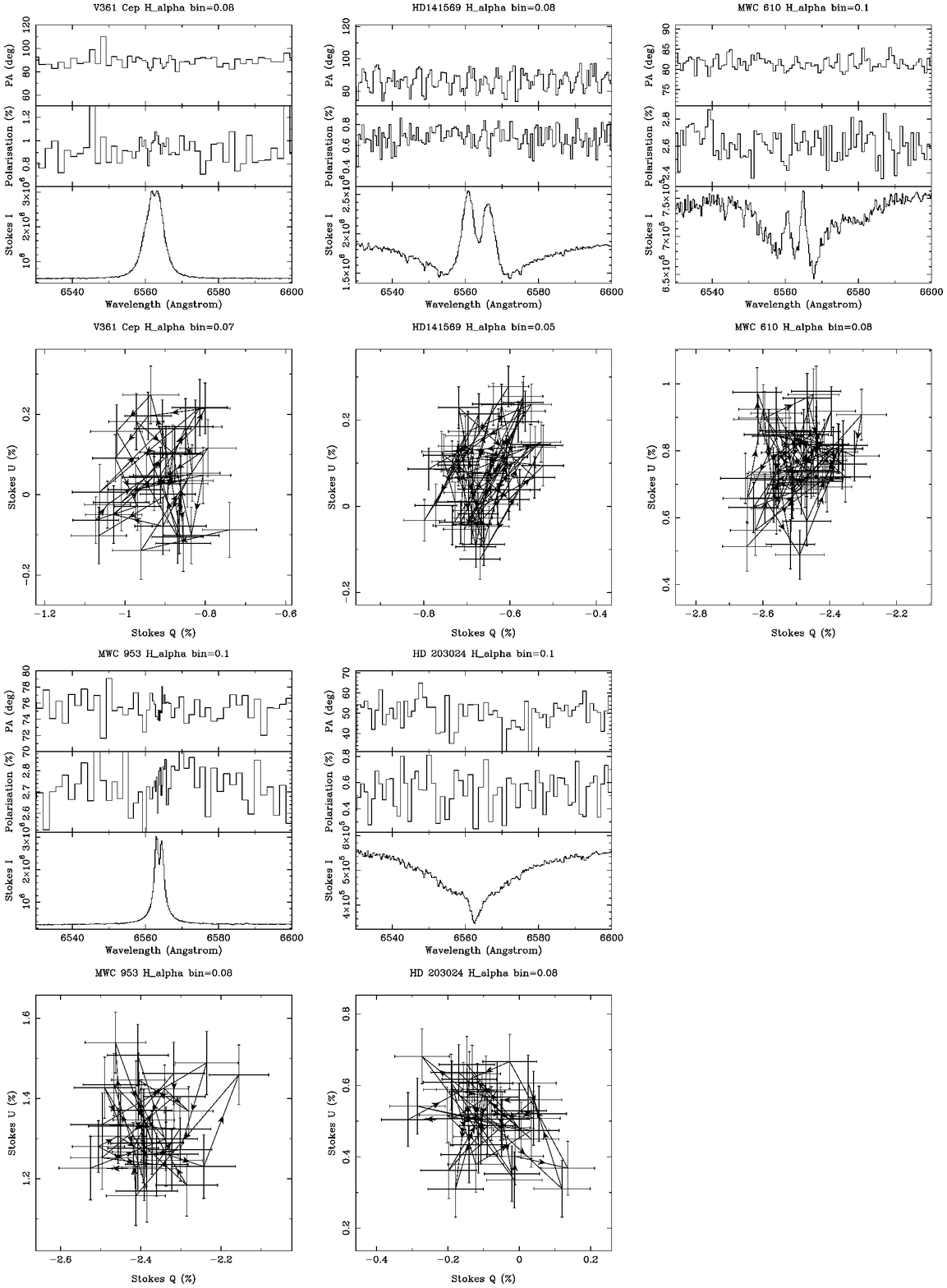}
       
   \caption{continued}

\end{figure*}



%%%%%%%%%%%%%%%%%%%%%%%%%%%%%%%%%%%%%%%%%%%%%%%%%%

% Don't change these lines
\bsp	% typesetting comment
\label{lastpage}
\end{document}